\documentclass{aa}  

\usepackage{graphicx}
\usepackage{txfonts}
\usepackage{natbib}
\bibpunct{(}{)}{;}{a}{}{,}
\begin{document} 

\title{A 4.6-year period brown-dwarf companion interacting with the hot-Jupiter CoRoT-20 b \thanks{Based on observations collected with the SOPHIE spectrograph on the 1.93-m telescope at Observatoire de Haute-Provence (CNRS), France, and with the HARPS spectrograph (Prog. 188.C-0779) at the 3.6-m telescope at La Silla Observatory.}}

\author{J. Rey\inst{1} \and F. Bouchy\inst{1} \and M. Stalport\inst{1} \and M. Deleuil\inst{2} \and G. H\'ebrard\inst{3,4} \and J. M. Almenara\inst{1} \and R. Alonso\inst{5,6} \and S. C. C. Barros\inst{16} \and A. Bonomo\inst{7}\and G. Cazalet\inst{2} \and J. B. Delisle\inst{1} \and R. F. D\'iaz\inst{8,9} \and M. Fridlund\inst{10,11} \and E. W. Guenther\inst{5,12} \and T. Guillot\inst{13} \and G. Montagnier\inst{3,4} \and C. Moutou\inst{14} \and C. Lovis\inst{1} \and D. Queloz\inst{15} \and A. Santerne\inst{2} \and S. Udry\inst{1}}

\institute{
Observatoire Astronomique de l'Universit\'e de Gen\`eve, 51 Chemin des Maillettes, 1290 Versoix, Switzerland 
\and 
Aix Marseille Universit\'e, CNRS, Laboratoire d'Astrophysique de Marseille UMR 7326, 13388 Marseille cedex 13, France 
\and 
Institut d'Astrophysique de Paris, UMR7095 CNRS, Universit\'e Pierre \& Marie Curie, 98bis boulevard Arago, 75014 Paris, France 
\and 
Observatoire de Haute Provence, CNRS, Aix Marseille Universit\'e, Institut Pyth\'eas UMS 3470, 04870 Saint-Michel-l’Observatoire, France 
\and 
Instituto de Astrofísica de Canarias, C. Via Lactea S/N, 38205, La Laguna, Tenerife, Spain
\and 
Universidad de La Laguna, Dept. de Astrofísica, 38206, La Laguna, Tenerife, Spain
\and
INAF – Osservatorio Astrofisico di Torino, via Osservatorio 20, I-10025 Pino Torinese, Italy 
\and 
Universidad de Buenos Aires, Facultad de Ciencias Exactas y Naturales. Buenos Aires, Argentina 
\and 
CONICET - Universidad de Buenos Aires. Instituto de Astronom\'ia y F\'isica del Espacio (IAFE). Buenos Aires, Argentina 
\and 
Department of Earth and Space Sciences, Chalmers University of Technology, Onsala Space Observatory, SE-439 92 Onsala, Sweden
\and 
Leiden Observatory, University of Leiden, PO Box 9513, NL-2300 RA Leiden, the Netherlands
\and
Thüringer Landessternwarte Tautenburg, Sternwarte 5, 07778, Tautenburg, Germany guenther@tls-tautenburg.de
\and
Université Côte d'Azur, OCA, CNRS, Nice, France
\and
CNRS, Canada-France-Hawaii Telescope Corporation, 65-1238 Mamalahoa Hwy., Kamuela, HI-96743, USA 
\and 
Astrophysics Group, Cavendish Laboratory, J.J. Thomson Avenue, Cambridge CB3 0HE, UK
\and
Instituto de Astrof\'isica e Ci\^encias do Espa\c{c}o, Universidade do Porto, CAUP, Rua das Estrelas, 4150-762 Porto, Portugal
}

\date{Received; accepted}

\abstract {We report the discovery of an additional substellar companion in the CoRoT-20 system based on six years of HARPS and SOPHIE radial velocity follow-up. CoRoT-20 c has a minimum mass of 17 $\pm$ 1 $M_{Jup}$ and it orbits the host star in 4.59$\pm 0.05$ years, with an orbital eccentricity of 0.60 $\pm$ 0.03. This is the first identified system with an eccentric hot Jupiter and an eccentric massive companion. The discovery of the latter might be an indication of the migration mechanism of the hot Jupiter, via Lidov-Kozai effect. We explore the parameter space to determine which configurations would trigger this type of interactions.}

\maketitle
%
\section{Introduction}
Since the discovery of the first hot Jupiter, 51 Peg b \citep{Mayor1995}, the formation and evolution of short-period massive planets has been a subject of debate. They were long thought to form beyond the ice line, followed by an inward migration \citep{Lin1996}, but it has been shown recently that they can also form in situ, via core accretion \citep[e.g.][]{Boley2016,Batygin2016}. In the first scenario, several migration processes are possible, like disk-driven migration \citep{Goldreich1980,Lin1986,Ward1997,Tanaka2002} or tidal migration \citep{Fabrycky2007,Wu2007,Chatterjee2008,Nagasawa2008}. 
Fingerprints of the different mechanisms can be found in the orbital characteristics of the hot Jupiters. The ones in eccentric and/or misaligned orbits are frequently presented as the result of multi-body migration mechanisms, like Lidov-Kozai \citep{Lidov1962,Kozai1962,Mazeh1979,Eggleton2001,Wu2003}, gravitational scattering \citep{Weidenschilling1996,Rasio1996, Ford2001} or secular migration \citep{Wu2011}. 
To determine how important is the role of multi-body migration in the production of hot Jupiters, a first step is to identify the perturbing body and constrain its orbital parameters. At least a dozen multiplanetary systems including a hot Jupiter and a massive companion with a fully-probed orbit have been identified by radial velocity and transit surveys \citep[e.g.][]{Wright2009, Damasso2015, Triaud2017}. Moreover, follow-up surveys have been carried out to specifically find these companions and estimate their occurrence rate. Three examples of this are the \textit{Friends of hot Jupiters} survey carried out at Keck with the HIRES spectrograph \citep{Knutson2014, Ngo2015, Piskorz2015, Ngo2016}, the long-term follow-up of WASP hot Jupiters with CORALIE \citep{Neveu2016}, and the GAPS programme with HARPS-N \citep{Bonomo2017}. The reported discoveries of an outer massive body in hot-Jupiter systems, especially those presenting high eccentricities, create ideal conditions for the Lidov-Kozai mechanism to occur. 
The exchange of angular momentum between the inner and outer bodies induces secular oscillations in the eccentricity and inclination of the inner planet, known as Lidov-Kozai cycles. At each phase of high eccentricity, strong tidal dissipation occurs in the inner planet when it passes through perihelion, resulting in an inward migration. As the planet gets closer to the star, the combined effect of tides in the star, its oblateness, and general relativity, counterbalances more efficiently the Kozai cycles, and their amplitude decreases towards the highest value of the eccentricity. Ultimately, these oscillations are annihilated. At that time, the planet will circularize under the action of tidal effects, and its semi-major axis will decrease at a higher rate \citep[see Fig.~1 from][]{Wu2003}. \\
However it is not evident if the identified companions actually play the role of perturbing the hot Jupiter. In fact, \cite{Knutson2014} did not find any statistically significant difference between the frequency of additional companions in systems with a circular and well-aligned hot Jupiter orbit, and those eccentric and/or misaligned. Other studies suggest that misalignments could also be primordial \citep{Spalding2015,Thies2011}, meaning that inclined hot Jupiters could also arise via disk-driven migration. Moreover, a statistical study performed on a sample of six hot Jupiters orbiting cool stars, with measured obliquities, and with outer companions identified \citep{Becker2017}, showed that these outer companions should typically orbit within 20-30 degrees of the plane that contains the hot Jupiter, suggesting that not many systems have the necessary architecture for processes such as Kozai-Lidov to operate. Therefore, it is important to know not only which fraction of hot Jupiters have a massive companion, but also which fraction of these companions are actually capable of triggering the migration.

CoRoT-20 is one of the planetary systems discovered by the CoRoT space mission \citep{Baglin2009,Moutou2013}. This system is composed of a 14.7-magnitude G-type star hosting a very high density transiting giant planet, CoRoT-20b \citep{Deleuil2012}, in an eccentric orbit of 9.24 days of period. It was identified thanks to three CoRoT photometric transits, and characterized with 15 radial velocity measurements using HARPS, FIES and SOPHIE spectrographs. In this paper, we report a new substellar companion orbiting CoRoT-20 thanks to six years of additional observations obtained with HARPS and SOPHIE spectrographs. CoRoT-20 is the first identified system with an eccentric hot Jupiter ($e \geq 0.2$) and an eccentric massive companion with a fully probed orbit. Therefore, it represents an excellent candidate to test tidal migration models. Because the mutual inclination of the two companions is unknown, we explore the parameter space to provide possible configurations that trigger the migration via Lidov-Kozai effect.

\section{Spectroscopic follow-up}
\begin{figure}[t!]
   \centering
   \includegraphics[width=\hsize]{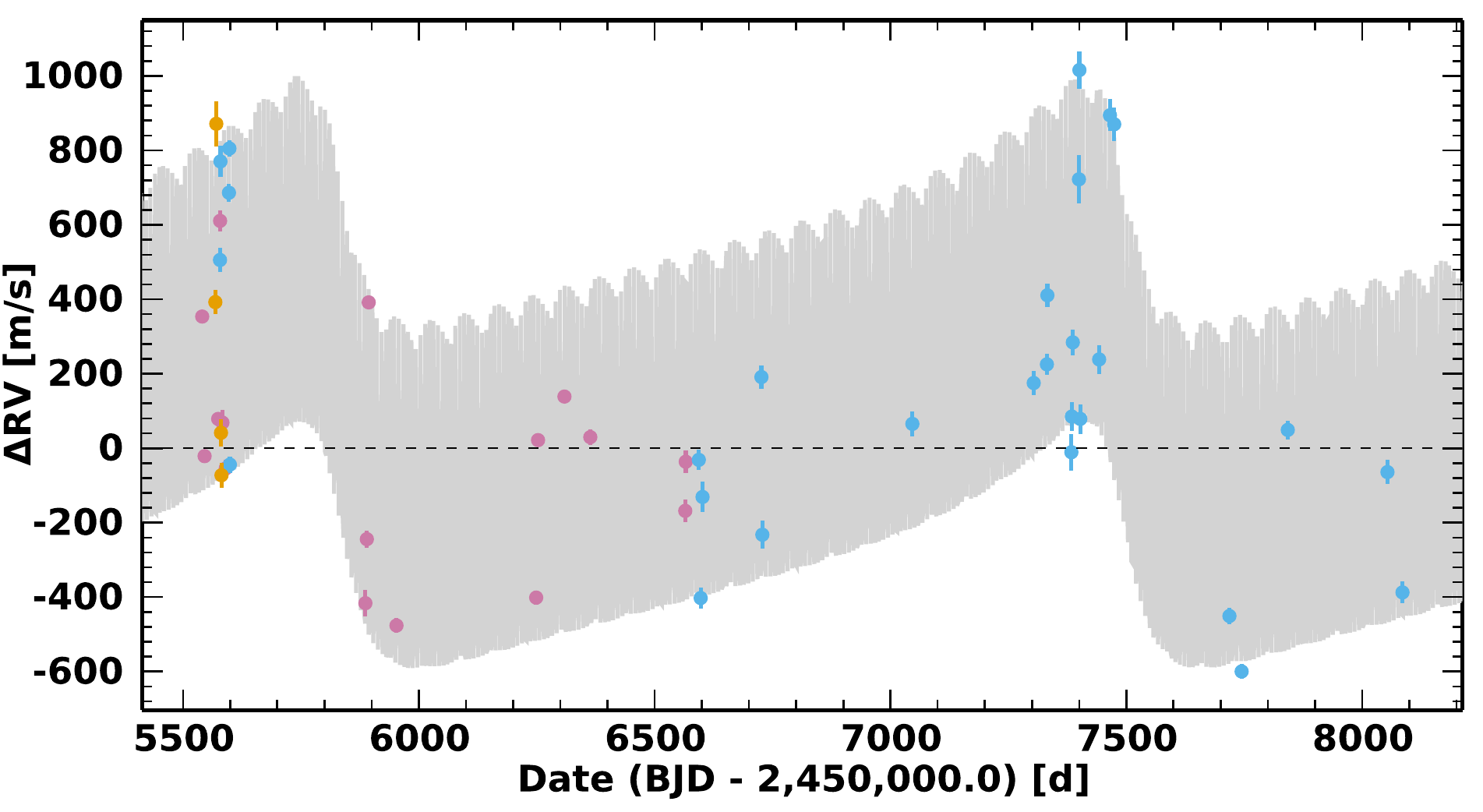}
   \includegraphics[width=\hsize]{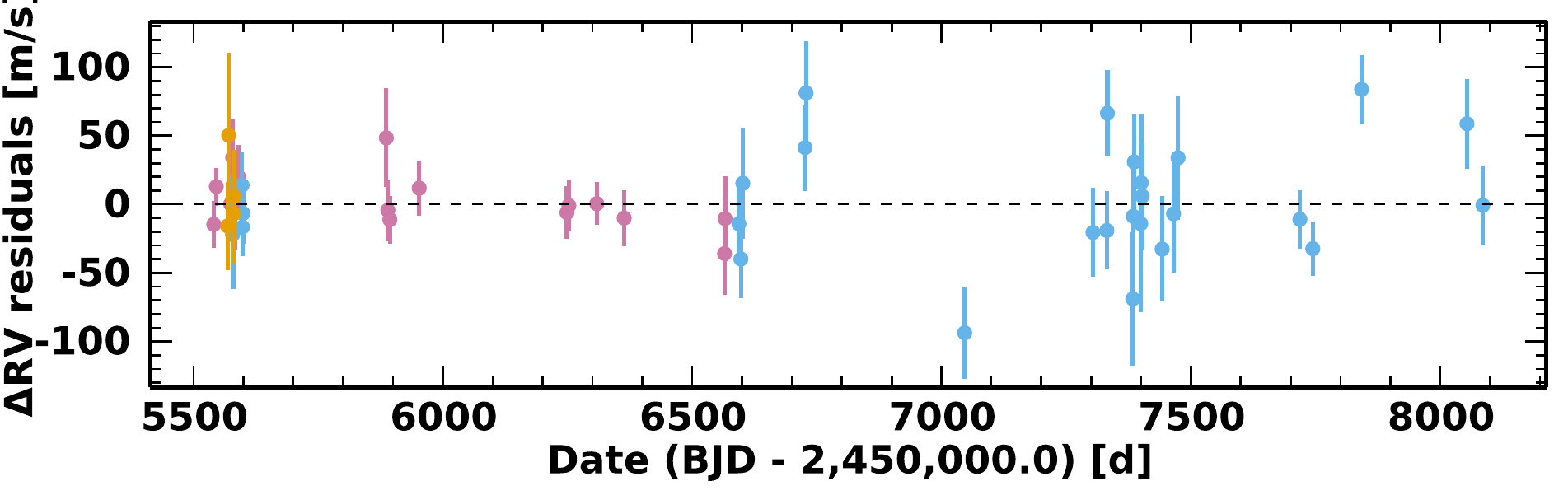}
   \includegraphics[width=\hsize]{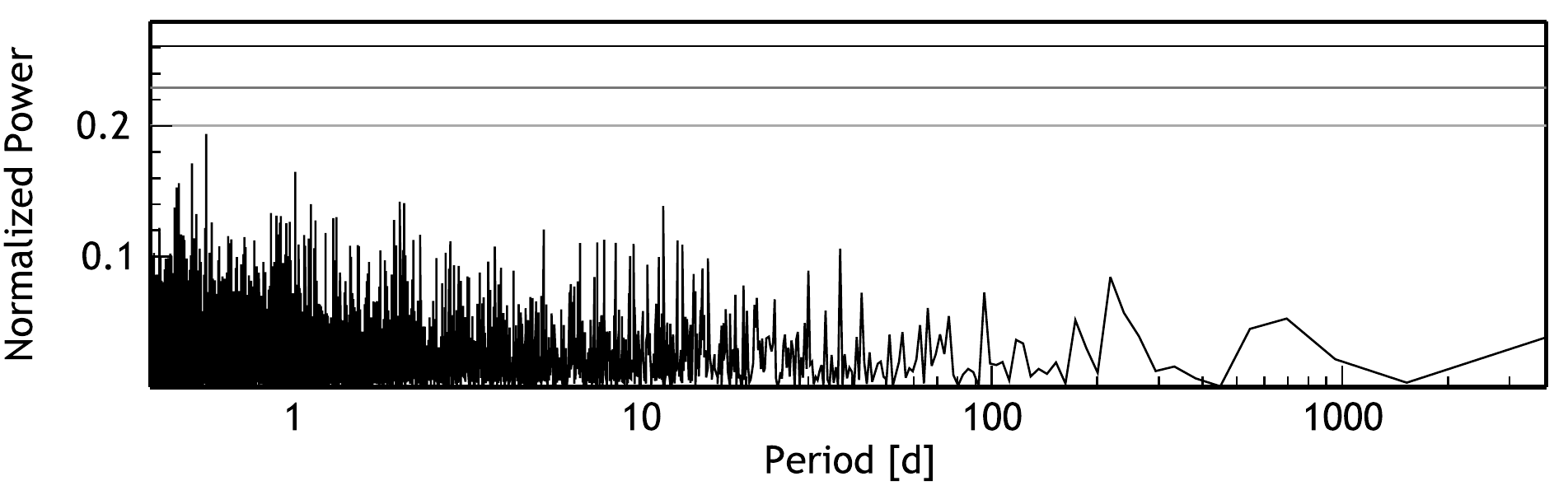}
      \caption{Radial velocity curve (top) and residuals (middle) of CoRoT-20 from FIES (orange), HARPS (purple) and SOPHIE (blue). Generalized Lomb-Scargle (GLS) periodogram (bottom) of the radial velocities after subtraction of the two orbits. False alarm probability levels are plotted for 50$\%$, 10$\%$ and 1$\%$.}
   \label{f:orbit}
\end{figure}

A long-term radial velocity monitoring of CoRoT-20 was done with HARPS spectrograph \citep{Mayor2003} from November 2011 to September 2013 and with SOPHIE spectrograph \citep{Perruchot2008,Bouchy2009} from October 2013 to November 2017. A total of 33 new radial velocity (RV) measurements spanning six years were obtained and are listed in Table \ref{t:rvs}. Previous observations from the discovery paper of CoRoT-20b \citep{Deleuil2012}, obtained between December 2010 and January 2011, are also included in our analysis. 
HARPS observing mode was exactly the same as described in \cite{Deleuil2012}. 

SOPHIE spectroscopic observations were done using the slow reading mode of the detector and high-efficiency (HE) {objAB} mode of the spectrograph, providing a spectral resolution of 39000 at 550nm, and where fiber B is used to monitor the sky background. 
The on-line data reduction pipeline was used to extract the spectra. The signal-to-noise (S/N) per pixel at 550 nm, obtained in 1-h exposures, is between 11 and 24. The radial velocities are derived by cross-correlating spectra with a numerical G2V mask \citep{Baranne1996,Pepe2002}. We also derive the FWHM, contrast and bisector span of the Cross-Correlation Function (CCF) as described by \cite{Queloz2001}. Some measurements contaminated by the Moon (flagged in Table \ref{t:rvs}), are corrected following the procedure described by \cite{Bonomo2010}. The charge transfer inefficiency (CTI) of the SOPHIE CCD, a systematic effect that affects the RVs at low S/N \citep{Bouchy2009b}, is also corrected following the empirical function described by \cite{Santerne2012}. Finally, the long-term instrumental instability was monitored thanks to the systematic observation in HE mode of the RV standard stars HD185144 and HD9407, known to be stable at the level of a few $ms^{-1}$ \citep{Bouchy2013}. We interpolate the RV variations of these standards and use it to correct our measurements following the procedure described by \cite{Courcol2015}. When no correction was possible, we quadratically added 13 ms$^{-1}$ to the uncertainties, that corresponds to the dispersion of the RV standard stars in HE mode \citep{Santerne2016}. Two SOPHIE spectra taken on 2013 March 25 and 26 were removed from our analysis due to a very low S/N and a strong contamination caused by the presence of the full Moon.

\section{Analysis and results}
\subsection{Orbit fitting with DACE}
For the orbital fitting and parameter determination, we used the Data and Analysis Center for Exoplanets (DACE\footnote{The DACE platform developed by the National Center of Competence in Research \textit{PlanetS} is available at \url{http://dace.unige.ch}}). DACE is a web platform dedicated to exoplanet data visualization and analysis. In particular, its tools for radial velocity analysis allow us to fit a preliminary solution by using the periodogram of our RVs. The analytical method used to estimate these parameters is described in \cite{Delisle2016}. We used this approach to fit a two-Keplerian model to the SOPHIE, HARPS and FIES data. For the hot Jupiter solution, the period and primary transit epoch are fixed to the values derived from the photometric analysis of \cite{Deleuil2012} and listed in Table \ref{t:param}. All other parameters (nine orbital parameters and the instrumental offsets) are let free.
The results of this preliminary solution are used as uniform priors for a Markov Chain Monte Carlo (MCMC) analysis, also available on DACE. The algorithm used for the MCMC is described in \cite{Diaz2014,Diaz2016}. The derived median parameters of the orbital solution are listed in Table \ref{t:param}. The error bars represent the $68.3\%$ confidence intervals.

\subsection{System parameters}
\begin{figure}
   \centering
   \includegraphics[width=\hsize]{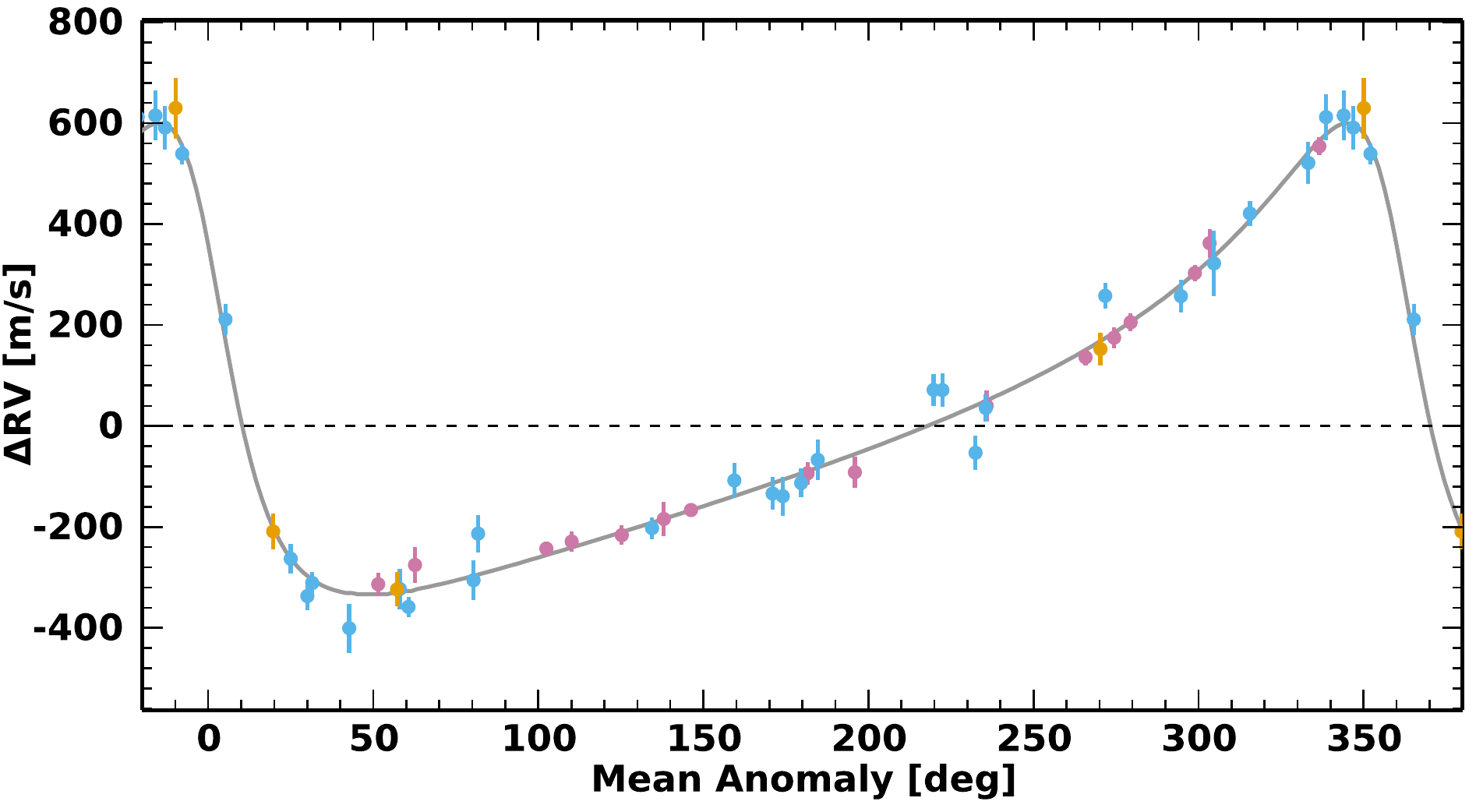}
   \includegraphics[width=\hsize]{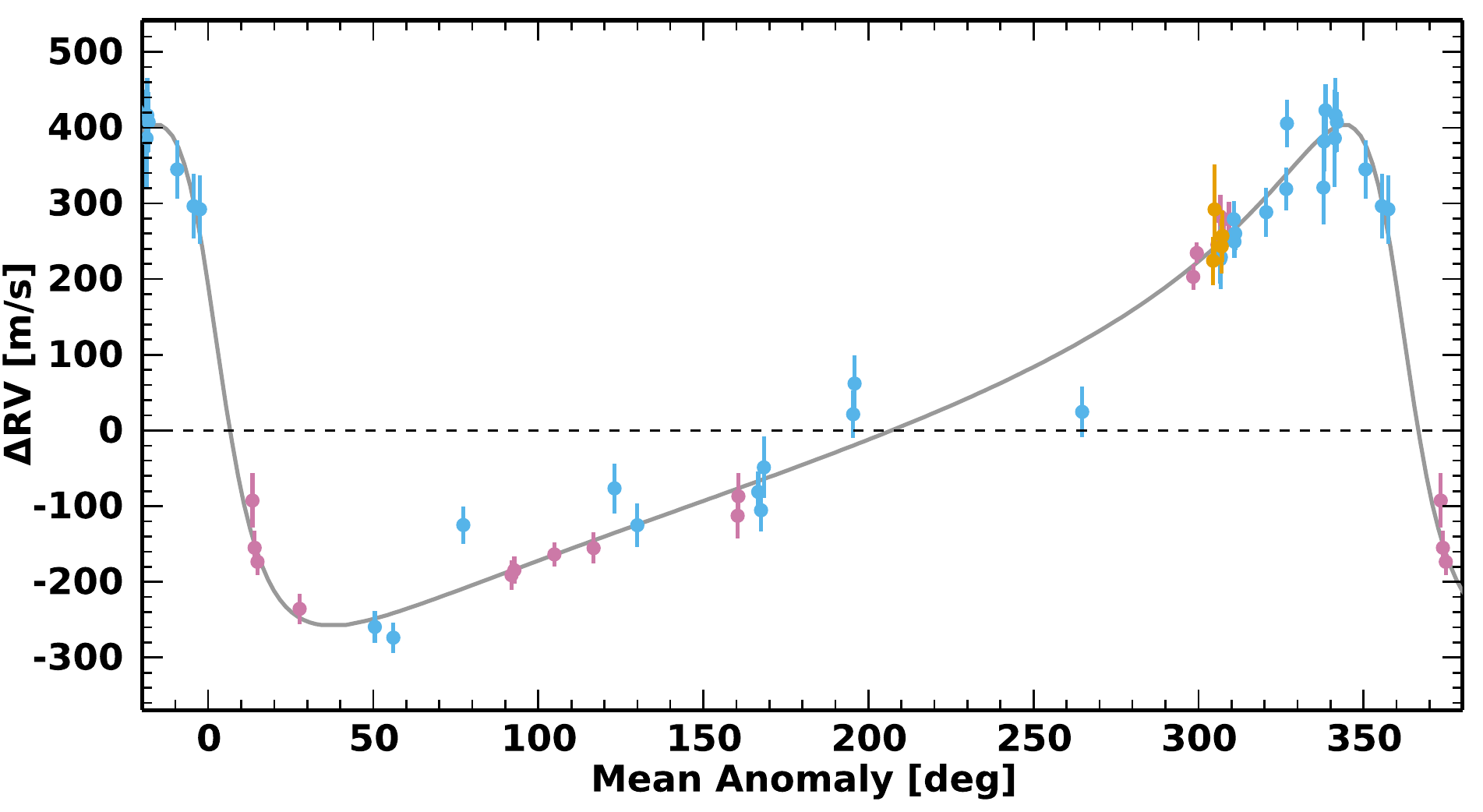}
      \caption{Phase folded radial velocities of CoRoT-20 b (top) and CoRoT-20 c (bottom).}
         \label{f:phase}
   \end{figure}

\begin{figure}
   \centering
   \includegraphics[width=\hsize]{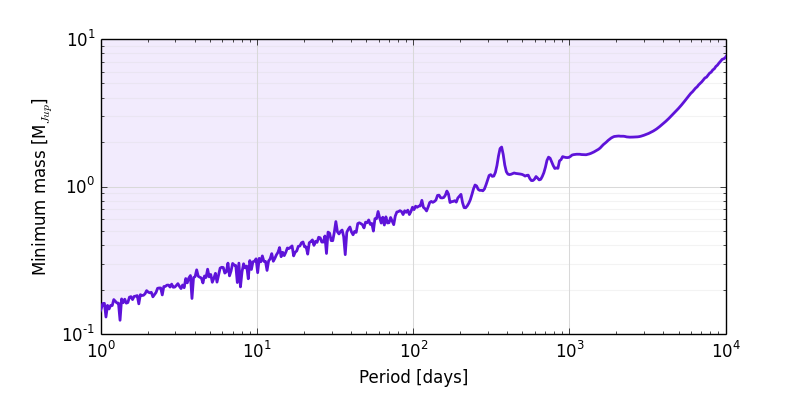}
      \caption{Mean mass limit detection of a third body on a circular orbit as a function of period, based on current radial velocity data residuals of CoRoT-20, after removing the two identified companions. We exclude the presence of any companions in circular orbits in the light-blue region.}
         \label{MassLimit}
   \end{figure}

Our long-term RV follow-up reveals the presence of an additional companion in the system, with a minimum mass of $m \sin i = 17$ M$_{Jup}$, orbiting the star in an eccentric orbit of 4.6 years. The best-fit orbit, residuals and periodogram of the residuals are shown in Fig. \ref{f:orbit}. Additionally, the phase folded radial velocities can be seen in Fig. \ref{f:phase}. It is noteworthy that both eccentricities and periastron arguments of the two orbiting companions are very similar. 
No correlations were found between the velocities and the CCF parameters, that could indicate the presence of stellar activity or blend with a stellar companion. Moreover, the photometric analysis of \cite{Deleuil2012} indicates that CoRoT-20 is a quiet star. No additional signals were found in the RV residuals. The parameters of CoRoT-20b are in agreement and within the error bars compared to those published by \cite{Deleuil2012}. Even though more radial velocities were added, there was no significant improvement in the precision of the hot Jupiter parameters. This is expected since we are including an additional orbit in our fit. To estimate the detection limits, we injected planets in circular orbits at different periods and phases to our RV residuals. These fake planets are considered detectable when the false alarm probability in the periodogram is equal or lower than $1\%$. The detection limits, shown in Fig. \ref{MassLimit}, allow us to exclude companions of 1 M$_{Jup}$ in orbits up to 100 days, and companions more massive than 10 M$_{Jup}$ in orbits up to 10,000 days (9.4 AU). CoRoT-20 was also part of the sample observed by \cite{Evans2016} using lucky imaging at the Danish 1.54-m telescope in La Silla. No physically associated stellar companions were found within 6". 

\begin{table*}
\caption{Stellar and orbital parameters}
\label{t:param}      
\centering                       
\begin{tabular}{l l l}       
\hline\hline             
\textbf{Ephemeris and stellar parameters}$^{(1)}$ & \textbf{Median values} & \textbf{Maximum of likelihood}\\
\hline  
Orbital period $P_b$ [days] & 9.24285 $\pm$ 0.00030 & ---\\
Primary transit epoch $T_{tr}$ [BJD] & 2455266.0001 $\pm$ 0.0014 & ---\\
Inclination $i_\textbf{b}$ [deg] & 88.21 $\pm$ 0.53 & ---\\
Stellar Mass $M_{\bigstar}$ [$M_{\odot}$] & 1.14 $\pm$ 0.08 & ---\\
\hline
\textbf{Planetary orbital parameters} & \textbf{Median values} & \textbf{Maximum of likelihood}\\
\hline
\textbf{Planet b} & & \\
RV semi-amplitude $K_b$ [m s$^{-1}$] & 467$^{+14}_{-13}$ & 470.93\\
Orbital eccentricity $e_b$ & 0.59 $\pm$ 0.02 & 0.58\\
Argument of periastron $\omega_b$ [deg] & 60.1$^{+2.5}_{-2.3}$ & 58.60\\
Orbital semi-major axis $a_b$ [AU] & 0.090 $\pm$ 0.002 & 0.09 \\
Mass $M_b$ [M$_{Jup}$] & 4.3 $\pm$ 0.2 & 4.25\\
\textbf{Companion c} & & \\
Orbital period $P_c$ [days] & 1675$^{+19}_{-17}$ & 1664.64\\
RV semi-amplitude $K_c$ [m s$^{-1}$] & 326$^{+19}_{-18}$ & 329.27\\
Orbital eccentricity $e_c$ & 0.60 $\pm$ 0.03 & 0.60\\
Argument of periastron $\omega_c$ [deg] & 65.0$^{+5.5}_{-5.7}$ & 66.82\\
Orbital semi-major axis $a_c$ [AU] & 2.90 $\pm$ 0.07 & 2.87\\
Minimum mass $M_c \sin i_\textbf{c}$ [M$_{Jup}$] & 17 $\pm$ 1 & 16.51\\
Periastron passage $T_p$ [BJD] & 2454136$^{+31}_{-35}$ & 2454152\\
\hline                                 
\end{tabular}
\begin{flushleft}
\begin{small}
$^{(1)}$ Parameters from \cite{Deleuil2012}
\end{small}
\end{flushleft}
\end{table*}

\section{Dynamical analysis with GENGA}

\begin{figure}
   \centering
   \includegraphics[width=\hsize]{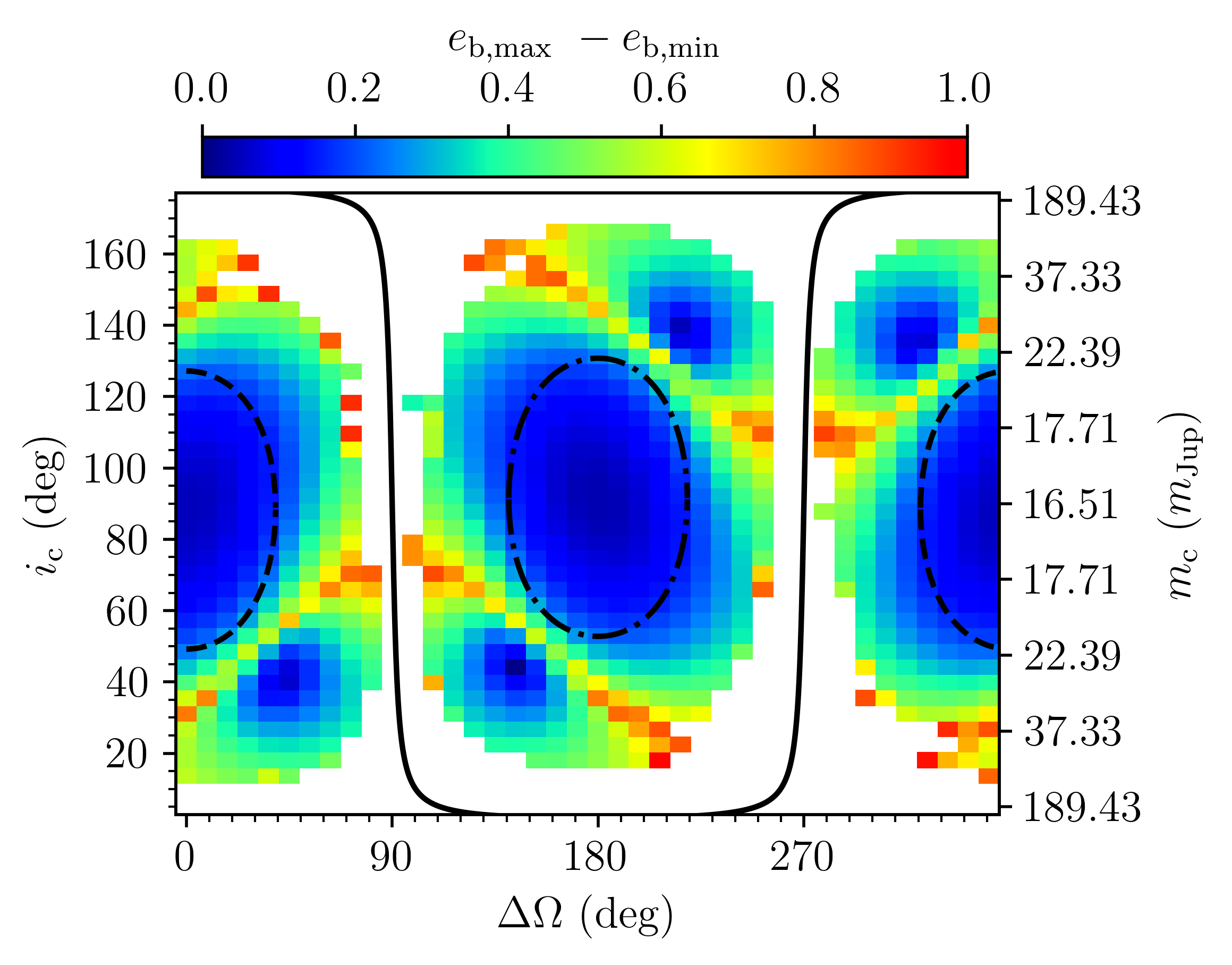}
      \caption{Difference between the maximum and minimum eccentricities of the inner planet over the whole simulation, for each set ($i_{c}$,$\Delta \Omega$). White squares represent aborted simulations (collision or ejection of one body). The black lines are isocurves of $I_m$, the mutual inclination. The solid line corresponds to $I_m = 90^{\circ}$, the dashed line to $I_m = 39^{\circ}$, and the dash-dotted line to $I_m = 141^{\circ}$.}
         \label{grid}
   \end{figure}

\begin{figure}
   \centering
   \includegraphics[width=\hsize]{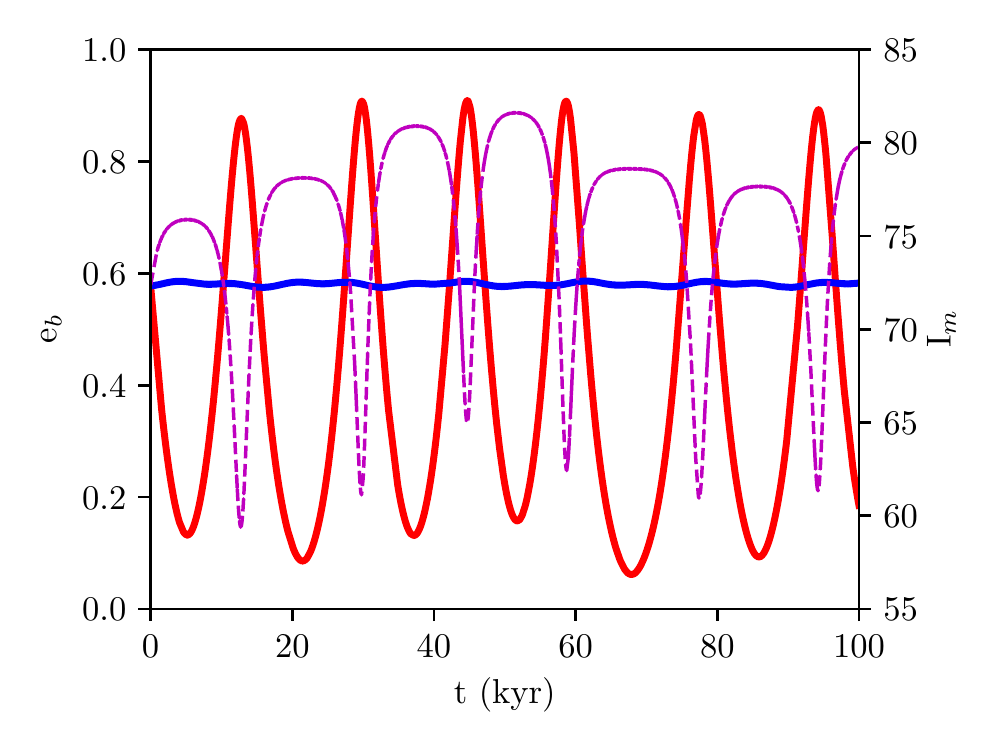}
      \caption{Temporal evolution of the eccentricity of the inner planet from initial conditions in both red and blue zones of Figure \ref{grid} (red and blue curves, respectively). The red curve corresponds to the values ($i_{c}$,$\Delta \Omega$) = ($70.4^{\circ}$,$72^{\circ}$) and an initial mutual inclination of $I_m = 72.5^{\circ}$. The evolution of the latter with time is represented by the dashed purple curve. The blue curve is associated to the initial set ($i_{c}$,$\Delta \Omega$) = ($44.2^{\circ}$,$144^{\circ}$) and $I_m = 122.8^{\circ}$. Its variation with time is negligible.}
         \label{e-vs-t}
   \end{figure}

\begin{figure}
   \centering
   \includegraphics[width=\hsize]{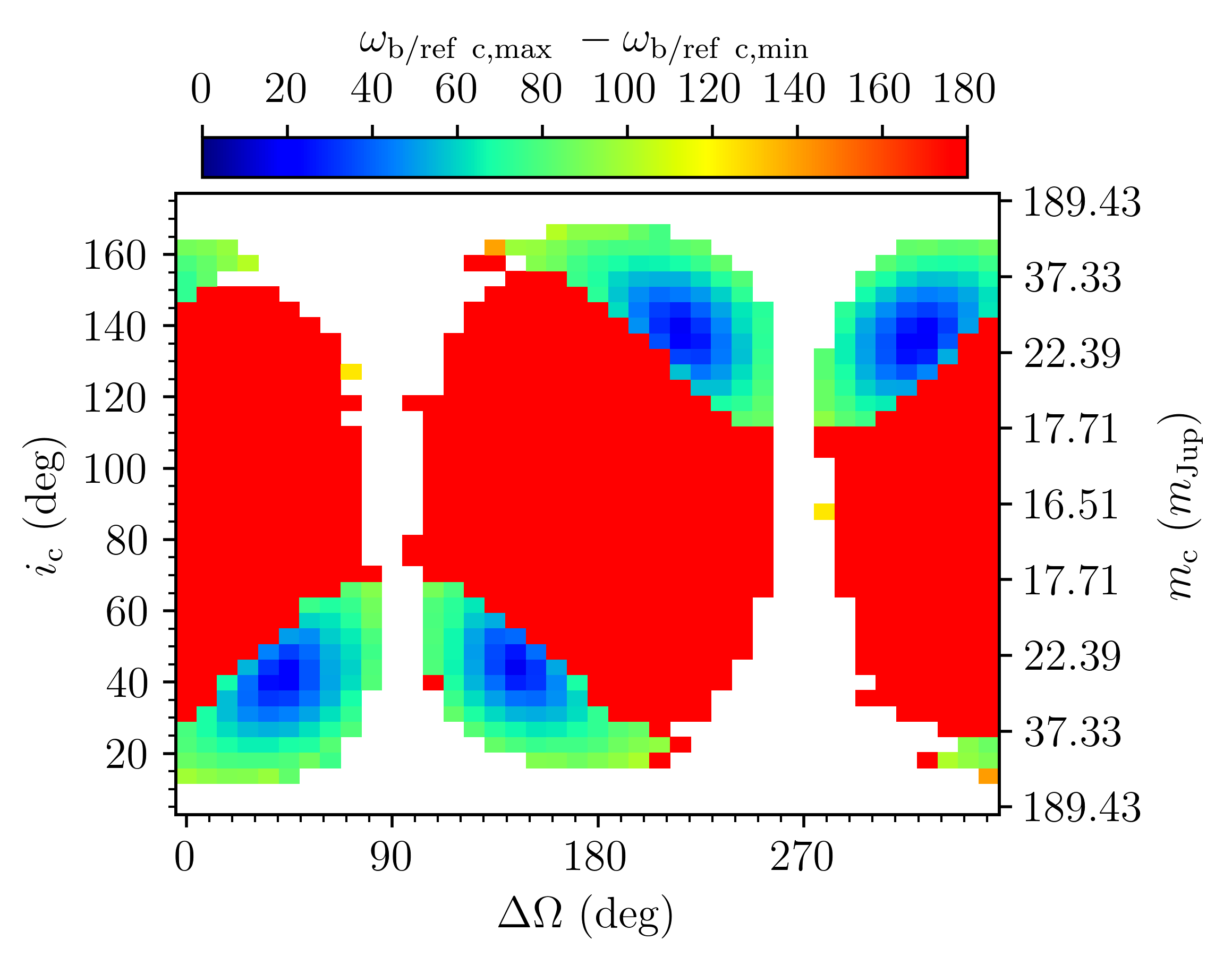}
      \caption{Amplitude of the variations of the argument of periastron of the inner planet $\omega_\mathrm{b/ref\ c}$, computed in the external body's reference frame for every initial set ($i_{c}$,$\Delta \Omega$). An amplitude of $180^{\circ}$ corresponds to a circulation of $\omega_\mathrm{b/ref\ c}$. Zones of libration exist for initially non coplanar configurations. These regions are identical to the low amplitude high eccentricity zones of Fig.~\ref{grid}.}
         \label{w1_2}
   \end{figure}

We performed numerical simulations using the GENGA integrator \citep{Grimm2014}. 
The initial orbital parameters of both companions were taken from the best fit values (maximum likelihood solution) of the MCMC analysis. The observations do not constrain the inclination of the outer body $i_{c}$ (defined in the same way as $i_b$, i.e. an orbit inclined of $90^{\circ}$ has its plane parallel to the line of sight), nor the relative longitude of the ascending nodes of the two bodies ($\Delta \Omega$ $\equiv$ $\Omega_{b}$ - $\Omega_{c}$) which has a dynamical influence. We thus explored these parameters on a 40x40 grid covering a large part of their domains ($i_{c} \in [5^{\circ};175^{\circ}]$ knowing that values near $0^{\circ}$ and $180^{\circ}$ are unstable because of perpendicular orbits between CoRoT-20~b and c, with an extremely large mass of the latter; $\Delta \Omega \in [0^{\circ};360^{\circ}]$). The known parameters were held fixed at their best-fit value over the grid (see Table \ref{t:param}), except the mass of the outer companion $M_{c}$. This was adjusted in accordance with $i_{c}$, $M_{c}\sin i_{c}$ being fixed by the radial velocity observations. Each simulation was integrated over $10^{5}$ years with a time step of 0.02 day, which is convenient with the perihelion passage of CoRoT-20~b. The General Relativity effects were included. In Fig.~\ref{grid}, we plot the results from the 1600 simulations as the maximum amplitude of the eccentricity oscillations of the inner planet. To this grid, we superimposed curves of fixed initial mutual inclination $I_m$ between CoRoT-20~b and c. The latter is defined as $\cos I_{m}~=~\cos i_{b}~\cos i_{c}~+~\cos \Delta\Omega ~\sin i_{b}~\sin i_{c}$, and therefore $I_m \in [0^{\circ};180^{\circ}]$. 
The dashed and dash-dotted curves delimit zones outside of which the mutual inclination is compatible with the appearance of Kozai cycles ($I_m \in [39^{\circ},141^{\circ}]$)\footnote{Strictly speaking, the critical values of $39^{\circ}$ and $141^{\circ}$ apply to the case of a circular external orbit and a massless inner body. However, these limits provide an adequate level of precision for a qualitative reasoning.}. Figure~\ref{grid} shows that such cycles might occur in the system. Based on analytical calculations from \citet{Fabrycky2007} and \citet{Matsumura2010}, we find that Kozai effect is the strongest in the CoRoT-20 system compared to general relativity and tides. Indeed, the time scales of the inner body's precession of argument of periastron are the following: due to the Kozai effect, $\tau_{K} \sim 2.3~10^{3} yr \sim 8.9~10^{4} P_{b}$; for general relativity, $\tau_{GR} \sim 4.2~10^{4} yr \sim 1.7~10^{6} P_{b}$; for tides\footnote{This estimate takes into account both the love numbers of the star and the inner planet. Arbitrary values for these were found in Table 1 from \citet{Wu2003}.}, $\tau_{tides} \sim 3.6~10^{5} yr \sim 14.3~10^{6} P_{b}$. \\ 
The largest Kozai cycles are located in the red zones, where the amplitude of the eccentricity oscillations is the most significant. 
To illustrate this, Fig.~\ref{e-vs-t} shows the temporal evolution of eccentricity of the inner planet for two different initial conditions corresponding to red and blue regions of Fig.~\ref{grid} (the red and blue curves, respectively). The purple dashed curve indicates the evolution of the mutual inclination associated to the same initial conditions as the red curve. Both lines being phase opposed, it clearly depicts an alternation between high eccentricity of the inner body and high mutual inclination, which is characteristic of Kozai cycles. Furthermore, the four small regions of low eccentricity variation of the inner planet from Fig.~\ref{grid} (two located at $i_c \sim 40^{\circ}$, and two at $i_c \sim 140^{\circ}$) are compatible with the Lidov-Kozai effect too. Indeed, these zones surround fixed points of high eccentricity and high mutual inclination of the phase space, as shown in Fig.~\ref{w1_2}. In this figure, the variation of the argument of periastron of the inner planet is calculated in the external body's frame, for each initial condition. The same small libration zones as in Fig.~\ref{grid} are observed, depicting an oscillation of both $e_b$ and $\omega_\mathrm{b/ref\ c}$ as expected nearby the Kozai fixed points \citep[see Fig. 5 from][]{Kozai1962}. By further comparing Fig.~\ref{grid} and \ref{w1_2}, we notice that the red boxes of Fig.~\ref{grid} are located in the circulation regime of $\omega_\mathrm{b/ref\ c}$. The corresponding Kozai cycles are thus qualified as rotating, while the small blue regions consist in librating type cycles.  \\ 
White zones from Fig.~\ref{grid} and \ref{w1_2} correspond to unstable regions of the parameter space, either because the inner planet collided onto the star (too high eccentricity) or because of a real instability (collision with the outer body, or ejection). We thereby exclude the corresponding doublets of parameters ($i_{c}$,$\Delta \Omega$). These sets maximize the mutual inclination $I_m$, i.e. they are associated to $I_m \sim 90^{\circ}$ and $I_m \sim 270^{\circ}$ (the isocurve of which is phase opposed to the former). However, let us remind that only the initial parameters $i_c$, $M_c$ and $\Delta \Omega$ were varied along the grid. All the others were fixed at their maximum of likelihood value. If we had changed them along the grid according to their posterior distributions, the results would probably have been slightly different as we would have explored different regions of the phase space. \\
  
The observations show a rather good alignment of the arguments of periastra (the best fit values exhibit $| \omega_b - \omega_c| \sim 8^{\circ}$). The nature of this alignment, coincidence or hidden dynamical process, could add constraints on the history of the system. We looked at the temporal evolution of this alignment in our simulations. In the observer's frame, $\omega_b - \omega_c$ actually librates in the non coplanar regions of the grid, and circulates elsewhere. The pattern is similar to Fig.\ref{w1_2}, except that the libration zones have a larger amplitude - between $80^{\circ}$ and $120^{\circ}$. 
The oscillations of $\omega_b - \omega_c$ indicate the existence of a dynamical process. The latter is naturally identified as Kozai cycles of librating type, as they impose an oscillation of $\omega_b$ while $\omega_c$ is nearly constant over time\footnote{CoRoT-20~c is at least four times more massive than CoRoT-20~b, and their period ratio $P_c /P_b \sim$ 180. Therefore, most of the angular momentum of the system comes from the outer body, i.e. the orbit of the latter is nearly a pure keplerian.}. 
However, due to the large amplitude of the oscillations of $\omega_b - \omega_c$, we investigated their significance. 
We found out that in the libration zones, the arguments of periastra spend approximately twice more time aligned ($| \omega_b - \omega_c| \leq 10^{\circ}$) than in the circulation regions, i.e. about 12$\%$ against 6$\%$\footnote{These proportions are actually valid for the beginning of the simulations. For some initial conditions, a slow increase of $|\omega_b - \omega_c|$ is superimposed to the oscillations, so that in the end only a small temporal fraction is spent in the alignment configuration.}. We thus interpret the similarity between $\omega_b$ and $\omega_c$ as a coincidence, with a probability of approximately 12$\%$ for observing it if the Lidov-Kozai effect is active. Indeed, the latter does not maintain a permanent alignment between the arguments of periastra of both bodies, because the oscillations of $\omega_b$ are too large in the observer's frame and the process does not lock $\omega_c$ at a fixed value.

\section{Discussion}
In single-planet systems, the circumstellar disk can induce moderate eccentricities on the body \citep[e.g.][]{Rosotti2017,Teyssandier2017}. However in multi-planet systems, the disk damps the eccentricities raised by the gravitational interactions between the different planets. Therefore, the currently high eccentricity of CoRoT-20~b is expected to be entirely due to the presence of CoRoT-20~c. There are at least three migration mechanisms that can explain the existence of close-in hot-Jupiters on eccentric orbits.
The Lidov-Kozai mechanism is one of them, and was studied in the previous section. Another scenario is the gravitational scattering. In this process, three or more massive bodies form around the central star and get unstable orbits. Due to close encounters, one of the bodies is ejected from the system, leaving the remaining planets on eccentric orbits. However, such a process hardly explains the existence of planets as close to the central star as CoRoT-20 b. The third mechanism is secular migration. In this scenario, a system composed of two or more well-spaced, eccentric and inclined planets with chaotic initial conditions will present an evolution that can lead to the existence of an eccentric hot Jupiter. Nevertheless, if such a process was taking place, we would expect the orbit of CoRoT-20~c to have evolved toward a nul eccentricity \citep{Wu2011}. Let us mention that \citet{Nagasawa2008} have explored the possibility of a coupling between the different mentioned mechanisms. Finally, \citet{Almenara2018} have recently discussed an alternative migration mechanism to explain eccentric hot Jupiters, based on interactions between two planets at low relative inclination. However, for this mechanism to work, it requires an oscillation of the angle $\varpi_{b} - \varpi_{c}$ over time, where $\varpi$ denotes the longitude of periastron ($\varpi = \Omega + \omega$). In other words, the coplanar system has to be located close to the high eccentricity fixed point of the phase space - see figure 12 from \cite{Almenara2018}. 
This is incompatible with our simulations, which show a circulation of $\varpi_{b}$ - $\varpi_{c}$ for $i_{b} \sim i_{c}$. 
Considering the system as described in this work, i.e. with two detected companions, the Lidov-Kozai mechanism seems to be the most likely and simple scenario to explain the current configuration. 
A recent paper from \citet{Wang2018} further consolidates this conclusion. They assert that the Lidov-Kozai mechanism is the one with the highest efficiency in producing hot Jupiters on eccentric orbits. From the results of our numerical simulations, we gave some constraints on the unknown parameters $\Delta \Omega$ and $i_{c}$. Based on the uncertainties on $i_{c}$, we derive the range of possible values for $M_{c}$. We find that $M_{c}$ is in the range 16.5 - 69.6 M$_{Jup}$, placing CoRoT-20~c in the domain of brown dwarfs.\\

Inferences about the formation of the system are highly speculative. However, our conclusions may initiate further investigations. We assert that the Lidov-Kozai migration may play a role in the actual state of CoRoT-20~b. If this is confirmed (by better constraints on $i_c$, $M_c$ or $\Delta \Omega$) , it would imply an outward formation of the planet followed by the said migration. Nevertheless, this is not incompatible with a formation relatively close to the star, i.e. inside the ice line, and at a small eccentricity. The high density of the planet - $8.87 \pm 1.10~ g/cm^{3}$ from \cite{Deleuil2012} - might be a clue to further study its formation. Concerning the external body, its high mass may be the result of a star-like formation, i.e. by gravitational collapse of a primordial nebula. The high eccentricity, and perhaps high inclination, observed would naturally result from this process as long as the circumstellar disk that existed around CoRoT-20 was sufficiently diffuse or short-lived to keep high values of these parameters. A measure of the spin-orbit misalignment between the star and CoRoT-20~b's orbit, via the Rossiter-McLaughlin effect, could clarify the value of the inclination of CoRoT-20~c \citep[e.g.][]{Batygin2012,Zanazzi2018}. The age of the star is presently poorly constrained ($T \in [60;900]$ Myr). Reducing this uncertainty might yield a clue in the formation process too. \\

\begin{figure}[!ht]
\centering
\includegraphics[width=0.5\textwidth]{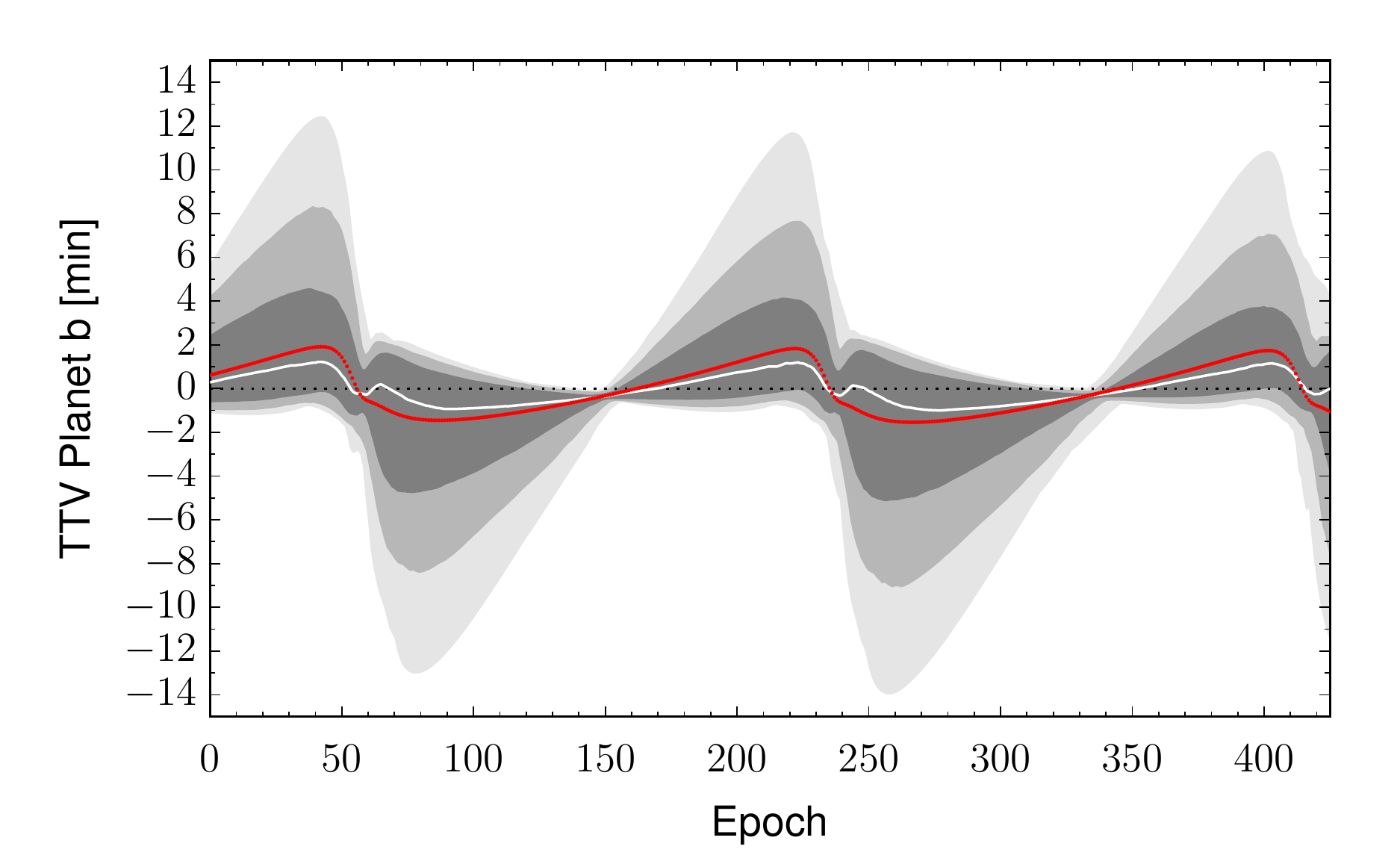}
\caption{Posterior TTVs of CoRoT-20b from the photodynamical modeling of the system. A thousand random draws from the posterior distribution are used to estimate the TTVs after the subtraction of linear ephemerids for each posterior sample. The three different gray regions represent the 68.3, 95.5, and 99.7\% credible intervals, the white line is the median value of the distribution, and the red line correspond to the TTVs of the maximum a posteriori model. The TTVs-amplitude is shown versus the CoRoT-20 b epoch number (0 is the first transit observed by CoRoT), up to the end of 2020.} \label{TTVs}
\end{figure}

Complementary observational techniques could potentially help constrain the mass of the second companion. We estimated the expected transit-timing variations (TTVs) of CoRoT-20b by performing the same photodynamical modeling as in \cite{Almenara2018}. We modeled the three transits observed by CoRoT \citep{Deleuil2012}, and the radial velocity measurements from HARPS, FIES and SOPHIE spectrographs. We used normal priors for the stellar mass and radius from \cite{Deleuil2012}, whereas non-informative uniform prior distributions were used for the rest of the parameters. The posterior TTVs of CoRoT-20b are plotted in Fig.~\ref{TTVs}. They have the periodicity of CoRoT-20c, and an amplitude < 5 min at 68\% credible interval. Furthermore, with this approach we can constrain the orbital inclination of CoRoT-20c to the range $[7, 172]^{o}$ at 95\% highest density interval, and its mass to 28$^{+35}_{-10}~M_J$ (68\% credible interval).

The \emph{Gaia} astrometric mission \citep{Gaia} will have a microarcsecond precision and a maximum of sensitivity close to the period of CoRoT-20~c. We expect an astrometric signature of at least $\alpha \sim33.6$ $\mu as$. With a total of 63 expected observations (from the \emph{Gaia} Observation Forecast Tool) and according to the magnitude of CoRoT-20, a final signal-to-noise of at least 3.8 would be obtained at the end of the mission. A combined analysis of radial velocity and astrometry would be challenging, but it would possibly help constrain the inclination and mass of the second body as well as the longitude of the ascending node.
Finally, if the system is close to coplanar, the second companion would have some probabilities to transit on mid-November 2020 but with an uncertainty of about 1 month. Assuming a radius of 0.8 R$_{Jup}$, the transit depth will be 6 mmag with a duration of up to 12 hours. This transit can potentially be detectable from dedicated ground-based photometric surveys like NGTS \citep{Wheatley2018}. When checking for previous transits (mid-October 2011 and end of March 2007), we see that the CoRoT observations of this system, done between March 1st and March 25th 2010, could not have covered it.
Measuring the stellar obliquity with the inner planet CoROT-20~b through the Rossiter- McLaughlin effect would provide additional constraints, as stated above. As explained in \cite{Deleuil2012} the expected amplitude of the RV anomaly of the Rossiter-McLaughlin effect (22 $\pm$ 5 m s$^{-1}$) is too small to be detected with HARPS according to the typical photon-noise uncertainty (20-30 m s$^{-1}$) but can be easily measured with ESPRESSO on the VLT \citep{Pepe2014}.

\begin{acknowledgements}
We gratefully acknowledge the Programme National de Plan\'etologie (telescope time attribution and financial support) of CNRS/INSU and the Swiss National Science Foundation for their support. We warmly thank the OHP staff for their support on the 1.93 m telescope. This work has been carried out in the frame of the National Centre for Competence in Research “PlanetS” supported by the Swiss National Science Foundation (SNSF). J.R. acknowledges support from CONICYT-Becas Chile (grant 72140583). SCCB acknowledges support from Funda\c{c}\=ao para a Ci\^encia e a Tecnologia (FCT) through national funds and by FEDER through COMPETE2020 by these grants UID/FIS/04434/2013 \& POCI-01-0145-FEDER-007672 and PTDC/FIS-AST/1526/2014 \& POCI-01-0145-FEDER-016886; and also acknowledges support from FCT through Investigador FCT contracts IF/01312/2014/CP1215/CT0004. Finally, we thank the referee for his/her thorough review and highly appreciate the comments and suggestions, which contributed to improving this publication.
\end{acknowledgements}

%
%

\bibliographystyle{aa}
\bibliography{corot20}

\begin{table*}
\caption{Log of additional radial velocity observations. Dates marked with an asterisk were corrected of moonlight contamination.}             
\label{t:rvs}      
\centering          
\begin{tabular}{l l l l l }     
\hline\hline       
Date & BJD-2450000 & RV & $\sigma$(RV) & Instrument\\
& [days] & [km s$^{-1}$] & [km s$^{-1}$] & \\
\hline                    
2011-11-21 & 5886.71941 & 59.958 & 0.036 & HARPS\\
2011-11-24 & 5889.77270 & 60.130 & 0.022 & HARPS \\
2011-11-28 & 5893.75303 & 60.766 & 0.017 & HARPS \\
2012-01-26 & 5952.63795 & 59.898 & 0.020 & HARPS \\
2012-11-17 & 6248.79863 & 59.973 & 0.019 & HARPS \\
2012-11-21 & 6252.75649 & 60.396 & 0.018 & HARPS \\
2013-01-16 & 6308.71329 & 60.513 & 0.016 & HARPS \\
2013-03-12 & 6363.54259 & 60.404 & 0.021 & HARPS \\
2013-09-29 & 6564.86859 & 60.206 & 0.030 & HARPS \\
2013-09-30 & 6565.89368 & 60.338 & 0.031 & HARPS \\
\hline    
2103-10-28$^{*}$ & 6593.61567 & 60.268 & 0.027 & SOPHIE \\
2013-11-01 & 6597.58122 & 59.897 & 0.028 & SOPHIE \\
2013-11-05 & 6601.55127 & 60.168 & 0.041 & SOPHIE \\
2014-03-09 & 6726.34430 & 60.490 & 0.031 & SOPHIE \\
2014-03-11 & 6728.30927 & 60.067 & 0.038 & SOPHIE \\
2015-01-23 & 7046.43549 & 60.365 & 0.033 & SOPHIE \\
2015-10-08$^{*}$ & 7303.65650 & 60.474 & 0.032 & SOPHIE \\
2015-11-05 & 7331.60716 & 60.525 & 0.029 & SOPHIE \\
2015-11-06$^{*}$ & 7332.63758 & 60.710 & 0.031 & SOPHIE \\
2015-12-27 & 7383.54897 & 60.289 & 0.049 & SOPHIE \\
2015-12-28 & 7384.51642 & 60.384 & 0.039 & SOPHIE \\
2015-12-30 & 7386.54467 & 60.584 & 0.034 & SOPHIE \\
2016-01-12 & 7399.51882 & 61.022 & 0.064 & SOPHIE \\
2016-01-13 & 7400.52700 & 61.315 & 0.050 & SOPHIE \\
2016-01-14 & 7402.42798 & 60.378 & 0.040 & SOPHIE \\
2016-02-23 & 7442.37874 & 60.538 & 0.038 & SOPHIE \\
2016-03-17 & 7465.30206 & 61.194 & 0.043 & SOPHIE \\
2016-03-26 & 7474.33253 & 61.170 & 0.045 & SOPHIE \\
2016-11-28 & 7718.64729 & 59.849 & 0.021 & SOPHIE \\
2016-12-21 & 7744.48132 & 59.700 & 0.020 & SOPHIE \\
2017-03-29 & 7842.32912 & 60.348 & 0.025 & SOPHIE \\
2017-10-27 & 8053.64794 & 60.235 & 0.033 & SOPHIE \\
2017-11-28 & 8085.55093 & 59.912 & 0.029 & SOPHIE \\
\hline
\end{tabular}
\end{table*}

\end{document}